\documentstyle[12pt]{article}

\newcommand{\be}{\begin{equation}}
\newcommand{\ee}{\end{equation}}
\newcommand{\sB}{\stackrel{\rightarrow}{B}}
\newcommand{\sS}{\stackrel{\rightarrow}{S}}
\newcommand{\sn}{\stackrel{\rightarrow}{n}}
\newcommand{\sr}{\stackrel{\rightarrow}{r}}
\newcommand{\sH}{\stackrel{\rightarrow}{H}}

\begin{document}

\begin{center}
{\large\bf{Relaxation Regimes of Spin Maser$^*$} \\ [5mm]
V.I.Yukalov} \\ [3mm]
{\it International Centre of Condensed Matter Physics \\
University of Brasilia, CP 04513 \\
Brasilia, DF 70919--970, Brazil \\
and \\
Bogolubov Laboratory of Theoretical Physics\\
Joint Institute for Nuclear Research\\
Dubna 141980, Russia}
\end{center}

\vspace{12cm}

$^*$ Report presented at the V International Workshop on Laser Physics 
(Moscow, July 22--26, 1996).

\newpage

\begin{abstract}

Spin relaxation in a microscopic model of spin maser is studied 
theoretically. Seven qualitatively different regimes are found: free 
induction, collective induction, free relaxation, collective relaxation, 
weak superradiance, pure superradiance, and triggered superradiance. The 
initiation of relaxation can be originated either by an imposed initial 
coherence or by local spin fluctuations due to nonsecular dipole 
interactions. The Nyquist noise of resonator does not influence processes 
in macroscopic samples. The relaxation regimes not initiated by an 
imposed coherence cannot be described by the standard Bloch equations.

\end{abstract}

\newpage

\section{Introduction}

Several experiments have been recently accomplished studying the 
peculiarities of spin relaxation in nonequilibrium systems of nuclear spins
coupled with a resonator. Such systems include $\;Al\;$ nuclear spins in 
ruby $\;(Al_2O_3)\;$ [1], and proton spins in propanediol 
$\;(C_3H_8O_2)\;$ [2--5], butanol $\;(C_4H_9OH)\;$, and ammonia 
$\;(NH_3)\;$ [6]. Radiation processes in these systems occur at 
radiofrequencies, that is, in the maser frequency region. Similar 
processes, also occurring at radiofrequencies, exist in ensembles of 
electron spins (see discussion and references in [7,8]). The main 
attention in studying these processes has been payed to the possibility 
of coherent effects in spin relaxation, the effects that resemble the well
known optical superradiance [9--11].

For describing spin relaxation, one usually invokes the Bloch equations 
supplemented by the Kirchhoff equation for resonator. In addition, to 
solve this complicated system of equations, one resorts to adiabatic 
approximation, when the resonator feedback field is proportional to 
transverse magnetization. The latter approximation is, of course, 
unphysical for transient phenomena and could be used only at the final 
stage of relaxation. Moreover, the Bloch equations as such, even being 
solved in a more realistic approximation [7], have the following generic 
limitations: (1) The role of dipole spin interactions is reduced to the 
inclusion of the spin--spin relaxation time $\;T_2\;$; (2) Coherent 
effects can be induced either by external magnetic field acting directly on 
spins or by imposing coherent initial conditions for transverse 
magnetization; (3) No self--organized coherence, that is called pure 
superradiance, can appear from a noncoherent state. These deficiencies 
do not  permit to reach good agreement with experiment. 

To give a realistic picture of spin dynamics in nuclear magnets, one has 
to deal with a microscopic model. This kind of model has been treated by 
means of computer simulations [12--14]. However, great number of 
parameters in the model makes it practically impossible to analyse 
different relaxation regimes by varying these parameters turn by turn. 
Here we present an analytical solution of the problem obtained with the 
help of a method developed in Refs. [15--17].

\section{Relaxation Regimes}

The microscopic model for a system of nuclear spins is given [18] by the 
Hamiltonian 
\be
\hat H = \frac{1}{2}\sum_{i\ne j}^NH_{ij} -\mu\sum_{i=1}^N\sB\sS_i
\ee
with dipole spin interactions
\be
H_{ij} = \frac{\mu^2}{r^3_{ij}}\left [\sS_i\sS_j - 3\left (\sS_i\sn_{ij}
\right )\left ( \sS_j\sn_{ij}\right )\right ] ,
\ee
where $\;\mu\;$ is a nuclear magneton and
$$ r_{ij} \equiv |\sr_{ij}|, \qquad \sr_{ij} \equiv \sr_i -\sr_j , \qquad
\sn_{ij} \equiv \frac{\sr_{ij}}{r_{ij}} . $$

The total magnetic field
\be
\stackrel{\rightarrow}{B} = \sH_0 +\sH
\ee
consists of a constant external field $\;\sH_0\;$ and an alternating 
field $\;H\;$ of a resonator coil,
\be
\sH_0 = H_0\stackrel{\rightarrow}{e}_z, \qquad \sH = 
H\stackrel{\rightarrow}{e}_x .
\ee

The equations of motion, to be considered in what follows, will involve
the average transverse spin component
\be
u \equiv \frac{1}{N}\sum_{i=1}^N\langle S_i^x - iS_i^y\rangle
\ee
and the average longitudinal component
\be
s \equiv \frac{1}{N}\sum_{i=1}^N\langle S_i^z\rangle ,
\ee
in which $\;\langle\ldots\rangle\;$ implies statistical averaging. Since (5)
is complex, we will need one more equation either for $\;u^*\;$ or for
\be
v \equiv |u| .
\ee
For the resonator magnetic field we introduce the dimensionless notation
\be
h \equiv \frac{\mu H}{\hbar\gamma_3} \qquad \left ( \gamma_3 \equiv 
\frac{\omega}{2Q}\right ) ,
\ee
in which $\;\gamma_3\;$ is the ringing width; $\;\omega\;$, the resonator 
natural frequency; and $\;Q\;$ is the resonator quality factor. The 
resonator is a coil of $\;n\;$ turns of a cross--section area $\;A_0\;$. For
the electromotive force $\;E_f\;$ it is convenient to pass to the 
dimensionless quantity
\be
f \equiv \frac{c\mu E_f}{nA_0\hbar\gamma_3^2} = f_0\cos\omega t .
\ee

For the functional variables (5)--(8), we can derive [16,17] the system 
of equations
\be
\frac{du}{dt} = i(\omega_0 - \xi_0 + i\gamma_2)u - i (\gamma_3h +\xi)s,
\ee
\be
\frac{ds}{dt} = \frac{i}{2}\left (\gamma_3h +\xi\right ) u^* -
\frac{i}{2}\left (\gamma_3h +\xi^*\right ) u - \gamma_1(s -s_\infty ) ,
\ee
\be
\frac{dv^2}{dt} = -2\gamma_2v^2 - i(\gamma_3h +\xi)su^* + 
i(\gamma_3h  + \xi^* )su ,
\ee
\be
\frac{dh}{dt} + 2\gamma_3h +\omega^2\int_0^t h(\tau)d\tau = -2\alpha_0
\frac{d}{dt}\left ( u^* + u\right ) + \gamma_3f ,
\ee
in which $\;\gamma_1\;$ is a spin--lattice width; $\;s_\infty\;$, a 
stationary magnetization per spin, and
$$ \omega_0 \equiv \frac{\mu H_0}{\hbar}, \qquad \gamma_2 \cong
\frac{\rho\mu^2}{\hbar} , $$
$$ \alpha_0 \equiv \pi\eta\frac{\rho\mu^2}{\hbar\gamma_3} \cong
\pi\eta\frac{\gamma_2}{\gamma_3} , $$
where $\;\rho\;$ is the density of spins and $\;\eta\;$, a filling 
factor. The set
$$ \bar\xi \equiv \{ \xi_0,\xi,\xi^*\} $$
of stochastic variables models local spin fluctuations, $\;\xi_0\;$ being 
responsible for secular dipole interactions while $\xi\;$ and $\;\xi^*\;$ 
for nonsecular dipole interactions. The distribution $\;p(\bar\xi)\;$ of 
these stochastic variables is assumed to be Gaussian with a width 
$\;\gamma_2^*\;$ corresponding to a nonhomogeneous width. Equations 
(10)--(13) are completed by the initial conditions
\be
u(0) =u_0, \qquad s(0) =z_0, \qquad v(0) = v_0, \qquad h(0) = 0.
\ee

To solve the system of nonliner integro--differential equations (10)-(13),
a method has been developed [15--17] combining the ideas of the 
multifrequency averaging method, guiding center approach, and of 
generalized asymptotic expansion. Here we shall delineate the main 
steps of the method only in brief.

\subsection{\it Classification of functional variables}

Take into account that there are the following small parameters
\be
\frac{\gamma_1}{\gamma_2} \ll 1, \qquad \frac{\gamma_2}{\omega_0} \ll 1,
\qquad \frac{\gamma_2^*}{\omega_0} \ll 1, \qquad 
\frac{\gamma_3}{\omega} \ll 1 .
\ee
Also, the quasi--resonance case will be considered, when the detuning is 
small:
\be
\frac{|\Delta|}{\omega_0} \ll 1 \qquad (\Delta\equiv\omega - \omega_0) .
\ee
As compared to the characteristic times
$$ T_1\equiv\frac{1}{\gamma_1}, \qquad T_2\equiv \frac{1}{\gamma_2}, \qquad
T_2^*\equiv \frac{1}{\gamma_2^*}, \qquad T_3\equiv \frac{1}{\gamma_3} , $$
the spin oscillation time
\be
T_0\equiv \frac{2\pi}{\omega_0} \ll \min\{ T_1,T_2,T_2^*,T_3\}
\ee
is small.

If all small parameters in (10)-(13) tend to zero, then
$$ \frac{du}{dt}\rightarrow \neq 0, \qquad \frac{ds}{dt}\rightarrow 0, $$
$$ \frac{dv^2}{dt}\rightarrow 0, \qquad
\frac{dh}{dt}\rightarrow \neq 0. $$
This shows that the variables $\;u\;$ and $\;h\;$ can be classified as 
fast and $\;s\;$ and $\;v^2\;$, as slow.

\subsection{\it Solution for fast variables}

The slow variables have the meaning of quasi--integrals of motion. Solve 
the equations (10) and (13) for the fast variables under fixed slow 
variables
$$ s=z, \qquad |u|^2 =v^2 , $$
treated as parameters in the corresponding solution
$$ u = u(z,v,t), \qquad h=h(z,v,t) . $$

\subsection{\it Temporal and stochastic averaging}

The right--hand sides of the equations (11) and (12) are to be averaged 
over the small oscillation time (17) and over stochastic local fields 
$\;\bar\xi\;$. This is to be done after substituting into these 
right--hand sides the solutions for the fast variables found at the 
previous step. The resulting equations
$$ \frac{dz}{dt} =\int\left ( 
\frac{1}{T_0}\int_0^{T_0} \frac{ds}{dt}dt\right ) p(\xi)d\xi , $$
$$ \frac{dv^2}{dt} =\int\left ( \frac{1}{T_0}\int_0^{T_0}
\frac{d|u|^2}{dt}dt\right ) p(\xi)d\xi $$
yield the solution
$$ z=z(t), \qquad v=v(t) . $$

\subsection{\it Definition of guiding centers}

The solutions $\;z(t)\;$ and $\;v(t)\;$ play the role of guiding centers 
for the slow variables. This role for the fast variables is played by
$$ \bar u = u(z(t),v(t),t) , $$
$$ \bar h = h(z(t),v(t),t) . $$

\subsection{\it Generalized asymptotic expansion}

Corrections to the guiding--center solutions can be obtained by expanding 
the functions 
$$ u = \bar u +\ldots , \qquad h =\bar h +\ldots , $$
$$ s=z+\ldots , \qquad |u|^2 = v^2 + \ldots $$
about the guiding centers and substituting these generalized asymptotic 
expansions into the equations (10)-(13). In the latter, one has to 
separate out corrections of different orders which leads to a system of 
linear equations.

This program has been accurately realized [16,17] yielding the following 
results. Considering the role of the Nyquist noise of resonator [19], it 
was found [20] that the relaxation parameter caused by the Nyquist noise 
is inversely proportional to the number of spins in the sample, 
$\;\gamma_N\sim N^{-1}\;$. The value of this parameter is negligibly 
small for macroscopic samples with $\;N\sim10^{23}\;$. 

The solutions for the slow variables read
\be
|u(t)|^2 =\left (\frac{\gamma_0}{g\gamma_2}\right ){\rm sech}^2\left (
\frac{t-t_0}{\tau_0}\right ) + 2\left ( \frac{\gamma_2^*}{\omega_0}\right )
s(t) , 
\ee
\be
s(t) =\frac{\gamma_0}{g\gamma_2}{\rm tanh}
\left (\frac{t-t_0}{\tau_0}\right ) - \frac{1}{g} .
\ee

Here
\be
g \equiv \alpha_0\left (\frac{\gamma_3}{\gamma_2}\right ) 
\frac{\pi(\gamma_2 -\gamma_3)^2}{(\gamma_2-\gamma_3)^2+\Delta^2}
\ee
is the effective parameter of coupling between the spin system and the 
resonator. Approximately,
$$ g \cong
\frac{\pi^2\eta(\gamma_2-\gamma_3)^2}{(\gamma_2-\gamma_3)^2+\Delta^2} . $$
Note that $\;g=0\;$ if $\;\gamma_2=\gamma_3\;$ and $\;\Delta\ne 0\;$.
The effective radiation width $\;\gamma_0\;$, related to the radiation 
time $\;\tau_0\;$ as $\;\gamma_0\tau_0=1\;$, is given by the equation
\be
\gamma_0^2 =\Gamma_0^2 +(g\gamma_2)^2\left [ v_0^2 - 2\left (
\frac{\gamma_2^*}{\omega_0}\right )^2z_0\right ] ,
\ee
in which $\;v_0=|u_0|,\; z_0=s(0)\;$, and
$$ \Gamma_0 =\gamma_2(1+gz_0) . $$
The delay time is
\be
t_0 = \frac{\tau_0}{2}\ln\left| 
\frac{\gamma_0 - \Gamma_0}{\gamma_0+\Gamma_0}\right | .
\ee

Qualitatively different regimes of relaxation can be classified according 
to the values of two parameters, $\;gz_0\;$ ans $\;gv_0\;$. 

\vspace{5mm}

\hspace{2cm} 1. {\it Free induction}:

$$ g|z_0| < 1 , \qquad 0 < gv_0 < 1 , $$
\be
t_0 < 0, \qquad \tau_0 \approx T_2 . 
\ee
This is the usual case of free nuclear induction, with the maximal 
coherence imposed at $\;t=0\;$.

\vspace{5mm}

\hspace{2cm} 2. {\it Collective induction}:

$$ gz_0 > -1 , \qquad gv_0 > 1 , $$
\be
t_0 < 0, \qquad \tau_0 < T_2 .
\ee
The coupling with the resonator is sufficiently strong to develop 
collective effects shortening the relaxation time $\;\tau_0\;$.

\vspace{5mm}

\hspace{2cm} 3. {\it Free relaxation}:

$$ g|z_0| < 1, \qquad v_0 = 0 , $$
\be
t_0 < 0 , \qquad \tau_0 \approx T_2 .
\ee
This corresponds to an incoherent relaxation due to the local fields.

\vspace{5mm}

\hspace{2cm} 4. {\it Collective relaxation}:
$$ gz_0 > 1 , \qquad v_0 = 0 , $$
\be
t_0 < 0 , \qquad \tau_0 < T_2 .
\ee
Collective effects shorten the relaxation time, as compared to the 
previous case.

\vspace{5mm}

\hspace{2cm} 5. {\it Weak superradiance}:

$$ -2 < gz_0 < -1 , \qquad v_0 = 0 , $$
\be
t_0 > 0, \qquad \tau_0 \approx T_2 .
\ee
Here the delay time is positive; the process is weakly coherent.

\vspace{5mm}

\hspace{2cm} 6. {\it Pure superradiance}:

$$ gz_0 < -2, \qquad v_0 =0. $$
\be
t_0 > 0 , \qquad \tau_0 < T_2 .
\ee
The coherence arises as a purely self--organized process started by local 
random fields and developed owing to the resonator feedback field.

\vspace{5mm}

\hspace{2cm} 7. {\it Triggered superradiance}:

$$ gz_0 < -1, \qquad gv_0 > 1 , $$
\be
t_0 > 0 , \qquad \tau_0 < T_2 .
\ee
This is a collective but not purely self--organized process, since the 
relaxation is triggered by an imposed initial coherence.

The term superradiance is used by analogy with the optical superradiance 
[9--11], although, as is discussed in Refs. [17,20], there are many 
principal differences between the atomic and spin superradiance.

The spin polarization, starting at $\;z_0=s(0)\;$, tends to
\be
s(t) \simeq \frac{\gamma_0}{g}\left ( T_2 -\tau_0\right ), \qquad t \gg t_0,
\ee
which shows that polarization reversal from negative $\;z_0\;$ to a 
positive $\;s(t)\;$ occurs when $\;\tau_0<T_2\;$.

To illustrate the numerical values of the parameters occurring in the 
report, let us write down some characteristic quantities typical of 
experiments with proton spin systems [2--6], such as propanediol 
($\;C_3H_8O_2\;$), butanol ($\;C_4H_9OH\;$), and ammonia ($\;NH_3\;$). 
The spin of a proton is $\;S=\frac{1}{2}\;$. The density of spins is 
$\;\rho\sim10^{22}cm^{-3}-10^{23}cm^{-3}\;$. The experiments are usually 
accomplished at low temperature $\;T\sim0.1K\;$ in a magnetic field 
$\;H_0\sim10^4G\;$. The Zeeman frequency is $\;\omega_0\sim10^8Hz\;$, and 
the corresponding wavelength is $\;\lambda\sim10^2cm\;$. The characteristic 
times are
$$ T_1\sim10^5sec, \qquad T_2\sim10^{-5}sec, \qquad T_3\sim10^{-6}sec-
10^{-5}sec. $$
The parameters in (15) are really small:
$$ \frac{\gamma_1}{\gamma_2} \sim 10^{-10}, \qquad  
\frac{\gamma_2}{\omega_0} \sim 10^{-3}, \qquad
\frac{\gamma_2^*}{\omega_0} \sim 10^{-3}, \qquad
\frac{\gamma_3}{\omega} \sim 10^{-2} . $$
The relaxation parameter $\;\gamma_N\;$ due to the resonator Nyquist 
noise, as compared to the spin--spin relaxation parameter $\;\gamma_2\;$, is
$$ \frac{\gamma_N}{\gamma_2}\sim 10^{-21}. $$

The picture obtained analytically is in agreement with the numerical 
simulations [12--14]. Some results of the latter are shown in Figs.1-4, 
where $\;K_{coh}\equiv P_{coh}/P_{inc}\;$ is a coherence coefficient being 
the ratio of the coherent part, $\;P_{coh}\;$, of the current power, 
$\;P\;$, to its incoherent part, $\;P_{inc}\;$, and $\;P_z\equiv 
-z(t)\;$. These figures illustrate the qualitative changes of the 
processes under the variation of some parameters. For qualitative 
understanding, the absolute values of units in figures are not important, 
so they are not specified (for details see [12--14]).

\section{Conclusion}

The main results of the analytical solution for the problem of spin 
relaxation in nuclear magnets can be formulated as follows.

\vspace{3mm}

1. Taking account of direct spin--dipole interactions is important for 
the correct description of relaxation processes.

\vspace{3mm}

2. Nonsecular dipole interactions play the major role in starting spin 
relaxation, when no initial coherence is thrust upon the system.

\vspace{3mm}

3. The Nyquist noise of a resonator does not influence the relaxation 
processes in macroscopic samples.

\vspace{3mm}

4. There are seven qualitatively different regimes of relaxation, three 
of which are triggered by primary coherence imposed upon the system and 
four others are initiated by local dipole fields.

\vspace{3mm}

5. The regimes that are purely self--organized, when no initial coherence 
is imposed, cannot be described by the standard Bloch equations. These 
regimes are: free relaxation, collective relaxation, weak superradiance, 
and pure superradiance.

\newpage

\begin{center}
{\bf Figure Captions}
\end{center}

{\bf Fig.1} 

\vspace{3mm}
Coherence coefficient $\;K_{coh}\;$, current power $\;P\;$, and the 
$\;z\;$--projection of spin polarization $\;P_z\;$ as functions of time, 
in arbitrary units, for two different coupling parameters, 
$\;g_1\;$ and $\;g_2\;$, with the relation $\;g_1/g_2=10\;$. The solid 
line is for $\;g_1\;$; the dashed, for $\;g_2\;$.

\vspace{1cm}

{\bf Fig.2}

\vspace{3mm}
The same as in Fig.1, for two different Zeeman frequencies, $\;\omega_{01}\;$
and $\;\omega_{02}\;$, related by the ratio $\;\omega_{01}/\omega_{02}=5\;$.
The solid line is for $\;\omega_{01}\;$; the dashed, for $\;\omega_{02}\;$.

\vspace{1cm}

{\bf Fig.3}

\vspace{3mm}
The same functions as in Fig.1, for different initial spin polarizations, 
with the relation $\;z_{01}/z_{02}=2\;$. The solid line is for $\;z_{01}\;$;
the dashed, for $\;z_{02}\;$.

\vspace{1cm}

{\bf Fig.4}

\vspace{3mm}
The same functions as in Fig.1, for different initial transverse 
magnetizations, with the relation $\;v_{01}/v_{02}=0.5\;$, when $\;z_0=0\;$.
The solid line is for $\;v_{01}\;$; the dashed, for $\;v_{02}\;$.

\end{document}